# Changes in Mean Global Physical Parameters of Blazhko RR Lyrae stars Derived from Multicolor Photometry

Á. Sódor

*Konkoly Observatory, P.O. Box 67, H-1525 Budapest, Hungary*
*email: sodor _at_ konkoly _dot_ hu*

**Abstract.** We developed an Inverse Photometric method (IPM) to determine global physical parameters of RR Lyrae stars exclusively from multicolor light curves (Sódor, Jurcsik & Szeidl, 2009, MNRAS, 394, 261). We showed that for good quality photometric observations of unmodulated RRab stars, the IPM gives similarly good results as direct Baade–Wesselink analyses do, but without the need for spectroscopic measurements. In the course of the development, we payed special attention to the applicability of the IPM for modulated RR Lyrae stars. Since there are no simultaneous spectroscopic radial velocity and photometric observations of any Blazhko star with good phase coverage both in pulsation and modulation, which would allow spectroscopic Baade–Wesselink analysis, the IPM is the only possibility today to study changes in the global physical parameters of Blazhko RR Lyrae stars during the modulation cycle. With the IPM, we have studied the extensive multicolor light curves of 4 Blazhko RRab stars to date. We observed these objects with the 24-inch telescope of the Konkoly Observatory during the past 5 years in the frame of the Konkoly Blazhko Survey. New results on two Blazhko stars (SS Cnc and RR Gem) are presented in this paper. Small but unambiguous changes in the pulsation-averaged mean temperature, mean radius and mean luminosity have been detected in both stars.

## INTRODUCTION

The traditional way of studying the global physical parameters ($T_{\text{eff}}$, $L$, $R$, $V_p$, $\log g$) and their changes with pulsation of RR Lyrae variables is to apply some form of the spectroscopic Baade–Wesselink (BW) analysis [1, 2, 3, 4, 5]. However, there are only about two dozens of field RR Lyrae stars and about one dozen such stars in globular clusters with published spectroscopic observations of radial velocities appropriate for BW analysis.

The Blazhko-effect, the phenomenon of light curve modulation shown by a significant fraction of RR Lyrae stars [6] is a more than hundred years old puzzle of stellar pulsation theory. The effect was discovered by Blazhko (1907) [7] and Shapley (1916) [8] but it does not have a satisfactory explanation yet.

To study the global physical parameters of Blazhko stars and their variations with both the pulsation and modulation cycles by spectroscopic BW analysis, an order of magnitude more radial velocity observations would be needed, with simultaneous photometric measurements, than available today on any such variable. Only this amount of data could ensure the appropriate coverage of both the pulsation and modulation cycles. The simultaneity of these observations is even more important than for unmodulated RR Lyrae stars, as in many cases the modulation is not stable on several years long timescales as it was shown e.g., for AH Cam [22], XZ Cyg [19], XZ Dra [20], RR Gem [16], AR Her [23], and RR Lyr [21].

There are many good quality multicolor light curves of fundamental mode Blazhko RR Lyrae stars available from the Konkoly Blazhko Survey [9, 6]. In order to extract more information from these data, we developed a new method to derive global physical parameters of RR Lyrae variables purely from their multicolor light curves, without the need for spectroscopic observations [10]. This Inverse Photometric Baade–Wesseling Method (IPM) was succesfully applied on two modulated RRab stars previously [MW Lyr – 11, DM Cyg – 12].

We have already published detailed analyses of the light curve modulation of the two RRab stars, SS Cnc and RR Gem, together with the multicolor photometric data of them, obtained during the Konkoly Blazhko Survey [SS Cnc – 13, RR Gem – 14, 16]. In this paper IPM results on these two stars are presented. The detected changes in the mean global physical parameters during the modulation cycle of all the four Blazhko stars studied so far (SS Cnc, DM Cyg, RR Gem, and MW Lyr) are shown here together.

## THE IP METHOD

From extended multicolor light curves of a Blazhko RR Lyrae star, the pulsation phase averaged mean $<V>$ magnitude and $<B-V>$ and $<V-I_C>$ colors can be determined for different phases of the modulation. These values, however, do not represent the variations in mean luminosity and mean temperature of the variable during the modulation cycle. To say the least, this is because different averages can be derived in different ways, e.g., either magnitude or intensity averages can be computed; calculating color indices might either precede averaging or the way around (see e.g. [13, 14, 26]). This is why we developed a more sophisticated method to derive global mean physical parameters for different phases of the modulation.

The IP method is capable for determining physical parameters of RRab variables and the variation of these parameters with pulsation phase ($\varphi$) exclusively from multicolor light curves [10]. This is an Inverse Photometric Baade-Wesselink analysis, which, using a nonlinear least squares algorithm, searches for the effective temperature [$T_{\rm eff}(\varphi)$] and pulsational velocity [$V_{\rm p}(\varphi)$] curves and other physical parameters that best fit the observed light curves, utilizing synthetic colors and bolometric corrections from static atmosphere models [27]. The initial $T_{\rm eff}(\varphi)$ and $V_{\rm p}(\varphi)$ curves are derived from empirical relations. The initial effective temperature is calculated from the input $(V-I_C)(\varphi)$ colour curve [24], while the initial pulsation velocity curve is derived either from the input $I_C(\varphi)$ light curve or from the template $V_{\rm p}(\varphi)$ curve of Liu (1991) [25]. The initial $T_{\rm eff}(\varphi)$ and $V_{\rm p}(\varphi)$ curves are then varied by the fitting algorithm.

The method yields the pulsation variations and absolute mean values of the radius ($R$), the effective temperature, the absolute visual brightness ($M_V$), and the luminosity ($L$) of individual objects. Mass ($m$) and distance ($d$) are also determined. Sometimes, however, the mass obtained by the IPM is too large, and a fixed value based on evolution theory has to be used instead. The distance can only be determined when the apparent magnitude zero points are known, which is sometimes not the case e.g., when differential photometry is available exclusively. Uncertainties of the IPM results are estimated by running the method with different internal settings. Details on the method and test results on non-modulated RRab stars can be found in Sódor et al. (2009) [10].

## THE DATA

Multicolor differential $BVI_C$ observations of the Konkoly Blazhko Survey [6], obtained with the 60 cm Heyde–Zeiss telescope of the Konkoly Observatory at Budapest, Hungary, are used for the present investigations. Light curve data of all the four objects have already been published [SS Cnc – 13, DM Cyg – 12, RR Gem – 14, 16, MW Lyr – 11]. Details on the instruments and data reduction procedure we applied can be found in Jurcsik et al. (2005) [14].

## IPM ANALYSIS

The data of MW Lyr and DM Cyg have already been analyzed with the IP method and the results have been published [11, 12]. Here these results are shown for comparison only.

## RR Gem

RR Gem has a short modulation period of 7.2 d with weak pulsation amplitude variations during the Blazhko cycle [14, 16]. The photometric data have been divided into 10 bins according to the Blazhko phase similarly to the division applied in the earlier analysis [14]. Light curves of these Blazhko phase bins have been analyzed by the IPM separately to determine the changes in the pulsation-phase-averaged mean global physical parameters with the modulation.

The parameters of RR Gem that do not change during the Blazhko cycle (mass – $m$, distance – $d$, reddening) have already been determined. We obtained mass and distance values using the mean light curve during the tests of the IPM [10]. These values are used here ($m = 0.81\ M_{\rm Sun}$, $d = 1240$ pc). The reddening of RR Gem we used was the one given by Liu & Janes (1990, Table 2) [17]. Mean values of the Blazhko phase dependent physical parameters are listed in Table 1. (Data from Sódor et al., 2009 [10].)

With the Blazhko phase independent parameters kept fixed at their values listed in Table 1, the IPM has been run for the multicolor pulsation light curves of each Blazhko phase.

**TABLE 1.** Pulsation and modulation averaged mean global physical parameters of RR Gem and SS Cnc. These values and their errors are calculated from the results obtained by running the IPM for the mean pulsation light curves of the Blazhko stars with different internal settings.

| Object (mass, [Fe/H]) | $<M_V>$ (mag) | $<L>$ ($L_{Sun}$) | $<T_{eff}>$ (K) | $<R>$ ($R_{Sun}$) | $<\log g_{stat}>$ |
|---|---|---|---|---|---|
| RR Gem (0.81, -0.15) | 0.66±0.01 | 47.5±1.0 | 6883±2 | 4.77±0.03 | 2.99±0.01 |
| SS Cnc (0.55, -0,12) | 0.70±0.01 | 46.1±0.5 | 6717±6 | 4.97±0.13 | 2.94±0.01 |

## SS Cnc

SS Cnc is another weakly modulated RRab star with short Blazhko period (5.3 d) [13]. Since the observations of this variable are somewhat sparser than that of RR Gem, a division of 7 Blazhko phase bins has been used for the IPM analysis to ensure complete pulsation phase coverage in each bin.

In the present analysis we accept a value of $m = 0.55\ m_{Sun}$ for the mass of SS Cnc. This mass value is consistent with evolution theory, considering the metal richness of SS Cnc; [Fe/H] = – 0.12, as derived from the Fourier parameters of the $V$ light curve using the empirical formula of Jurcsik & Kovács (1996) [18].

The intrinsic (dereddened) mean colors of SS Cnc can be determined by applying the IPM on the mean pulsation light curves fitted to the whole data sets of the 3 photometric bands. As a result we obtained $(B – V)_0 = 0.37$ mag and $(V – I_C)_0 = 0.45$ mag values. The mean values of other, Blazhko phase dependent parameters have also been determined during these runs and are listed in Table 1. Since there is no photometric standard star in the field of SS Cnc, the apparent $BVI_C$ magnitude zero points are not known and the distance cannot be determined with sufficient precision by the IPM. However, the uncertainty of the distance does not affect any of the other physical parameters, the IPM is capable to determine.

In order to derive the changes of the pulsation averaged mean global physical parameters with the Blazhko cycle, the IP method has been run for the 7 different Blazhko phases, after these parameters were determined and fixed.

Similarly to the Blazhko stars analyzed earlier (MW Lyr and DM Cyg) we have found that the actual values of the Blazhko phase independent parameters of RR Gem and SS Cnc influence only the mean values of those parameters that do vary with the Blazhko cycle. The shape and amplitude of their variations with Blazhko phase around these mean values practically do not depend on the actual values of the fixed parameters.

## RESULTS

Results of the IPM analysis on the four Blazhko stars are summarized in Fig. 1. Top panels show the $V$ light curves folded with the pulsation periods. The second row of the panels characterize the amplitude modulation with the amplitude of the $f_0$ pulsation component in the different Blazhko phase bins in $V$ band. The third row in Fig. 1 shows the phase modulation, interpreted as modulation in the pulsation period, as derived from the phase variations of the $f_0$ pulsation component corresponding to a fixed epoch in each Blazhko phase bin (see details in [11]). Bottom 3 rows of the panels show the variations in pulsation phase averaged temperature, radius and luminosity, respectively. All the panels show relative variations of the quantities with respect to their Blazhko phase averaged values.

In each Blazhko phase, results obtained with different internal settings of the IPM are shown. The spread of these points indicate the uncertainty of the results (Sódor et al., 2009 [10]).

The IPM solutions display slight but unambiguous changes in all the investigated global physical parameters of all the four stars. Not surprisingly, stronger modulation is accompanied with stronger variations in these parameters.

We note that the derived mean temperature, mean radius and mean luminosity are not expected to fulfill the Stefan–Boltzmann (SB) law in any Blazhko phase. The IPM solution ensures that these quantities fulfill the SB law in each phase of the pulsation. Since the SB law is non-linear and we calculate simple arithmetic averages according to the pulsation cycle to determine the pulsation phase averaged values in each Blazhko phase, the SB law does not apply for these averages anymore.

## DISCUSSION

We presented new IPM analyses of RR Gem and SS Cnc, and compared them with the results of the already analyzed MW Lyr and DM Cyg. We revealed definite changes in all the investigated global physical parameters in all four objects with amplitudes of the order of 0.1 – 4 %.

Since there are no appropriate spectroscopic radial velocity and photometric observations of any Blazhko star with good phase coverage both in pulsation and modulation, which would allow spectroscopic BW analysis, the IPM is the only possibility today to study changes in global physical parameters of Blazhko RR Lyrae stars during the modulation cycle.

To draw firm conclusion on the possible relations between the variations of the physical parameters, including their phases and amplitudes, and the modulation properties of RR Lyrae stars, i.e., the phase and amplitude modulation and the phase difference between these two, further IPM analyses of more Blazhko stars are needed. We observed several Blazhko RR Lyrae stars during the Konkoly Blazhko Survey to date [6]. Most of the multicolor Blazhko observations are extended enough to allow IPM analysis. We continue to process these data with the IP method to find these possible relations. This might help us to reveal how the global physical parameters are affected by the Blazhko effect and to get closer to the solution of this century-old enigma of RR Lyrae stars.

## ACKNOWLEDGMENTS

The financial support of OTKA grant T-068626 is acknowledged.

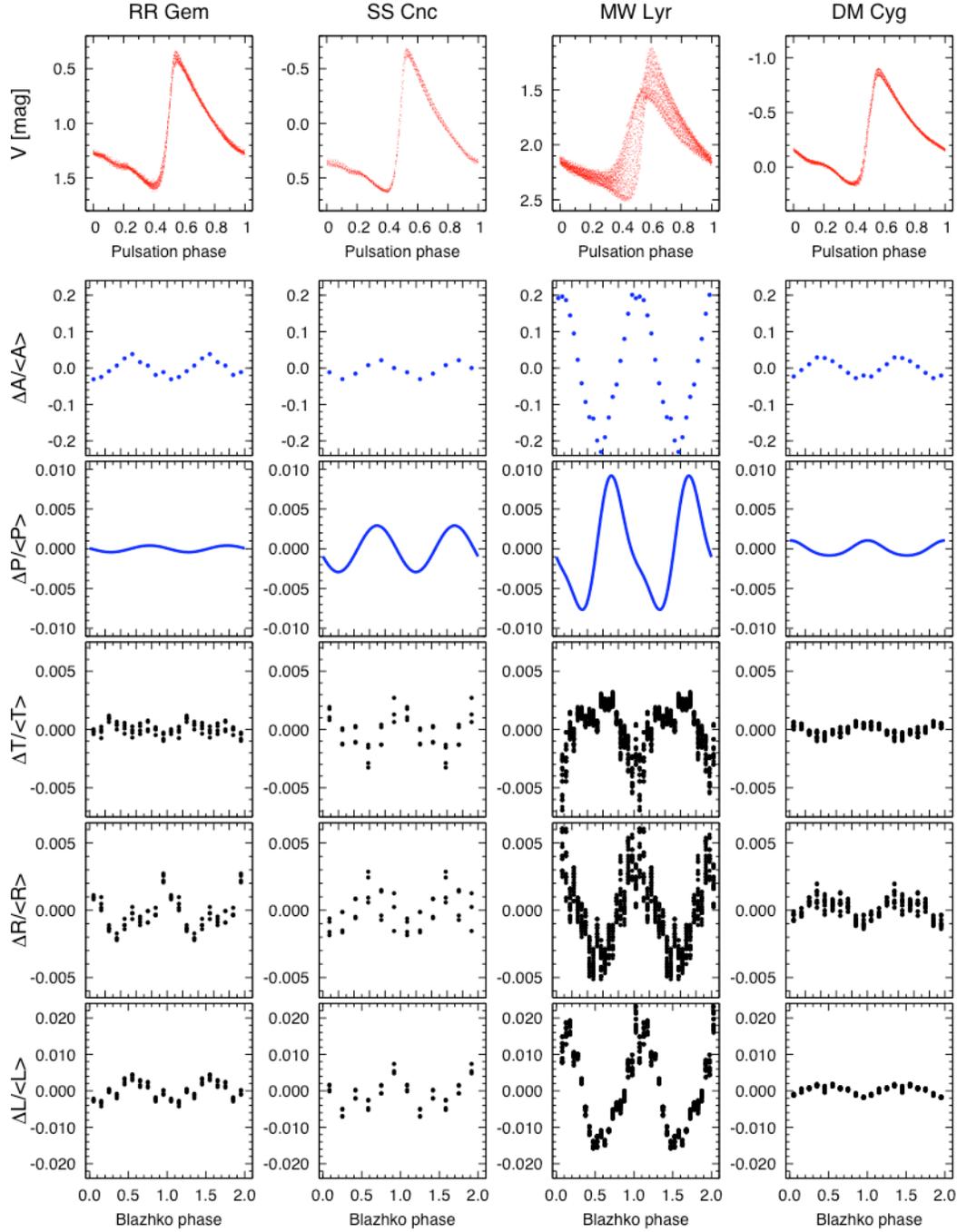

**FIGURE 1.** Top panels show the $V$ light curves folded with the pulsation period. The second row of the panels characterize the amplitude modulation with the amplitude of the $f_0$ pulsation component in the different Blazhko phase bins in $V$ band. The third row shows the phase modulation, interpreted as modulation in the pulsation period, as derived from the phase variations of the $f_0$ pulsation component corresponding to a fixed epoch in each Blazhko phase bin. Bottom 3 rows of the panels show the variations in pulsation phase averaged temperature, radius and luminosity, respectively. All the panels show relative variations of the quantities with respect to their Blazhko phase averaged values.